\documentclass[useAMS,usenatbib]{mn2e}
\usepackage{epsfig}
\usepackage{amssymb}
\usepackage{amsmath}

\title[The Most Magnetic Stars] 
{The Most Magnetic Stars}

\author[Dayal T. Wickramasinghe, Christopher A. Tout \& L. Ferrario]
{Dayal T. Wickramasinghe$^1$, Christopher A. Tout$^{1,2,3}$ and Lilia Ferrario$^1$\\
$^1$Mathematical Sciences Institute, The Australian National University,
ACT 0200, Australia\\
$^2$Institute of Astronomy, The Observatories, Madingley Road,
Cambridge CB3 0HA\\
$^3$Monash Centre for Astrophysics, School of Mathematical Sciences,
Building~28, Monash University, Victoria 3800, Australia}

\begin{document}

\date{Accepted.  Received ; in original form} 
\pagerange{\pageref{firstpage}--\pageref{lastpage}} \pubyear{}

\maketitle

\label{firstpage}

\begin{abstract}

Observations of magnetic A, B~and O~stars show that the poloidal
magnetic flux per unit mass $\Phi_{\rm p}/M$ appears to have an upper
bound of approximately $10^{-6.5}\,\rm G\,cm^2\,g^{-1}$.  A similar
upper bound to the total flux per unit mass is found for the magnetic
white dwarfs even though the highest magnetic field strengths at their
surfaces are much larger.  For magnetic A and~B stars there also
appears to be a well defined lower bound below which the incidence of
magnetism declines rapidly.  According to recent hypotheses, both
groups of stars may result from merging stars and owe their strong
magnetism to fields generated by a dynamo mechanism as they merge.  We
postulate a simple dynamo that generates magnetic field from
differential rotation.  We limit the growth of magnetic fields by the
requirement that the poloidal field stabilizes the toroidal and vice
versa.  While magnetic torques dissipate the differential rotation,
toroidal field is generated from poloidal by an $\Omega$ dynamo.  We
further suppose that mechanisms that lead to the decay of toroidal
field lead to the generation of poloidal.  Both poloidal and toroidal
fields reach a stable configuration which is independent of the size
of small initial seed fields but proportional to the initial
differential rotation.  We pose the hypothesis that strongly magnetic
stars form from the merging of two stellar objects.  The highest
fields are generated when the merge introduces differential rotation that
amounts to critical break up velocity within the condensed object.
Calibration of a simplistic dynamo model with the observed maximum
flux per unit mass for main-sequence stars and white dwarfs indicates
that about $1.5\times 10^{-4}$ of the decaying toroidal flux must
appear as poloidal.  The highest fields in single white dwarfs are
generated when two degenerate cores merge inside a common envelope or
when two white dwarfs merge by gravitational-radiation angular
momentum loss.  Magnetars are the most magnetic neutron stars.  Though
these are expected to form directly from single stars, their magnetic
flux to mass ratio indicates that a similar dynamo, driven by
differential rotation acquired at their birth, may also be the source
of their strong magnetism.

\end{abstract}


\section{Introduction}

Until recently, observations of magnetism in mid- to early-type stars
have focused on the chemically peculiar A~and late-B stars.  These
studies, which are biased towards the stronger field objects, have
revealed that there is an upper limit to the dipolar magnetic field
emerging from the stellar surface of $B_{\rm p} \approx 30\,$kG
\citep{elkin2010} for stars in the mass range $1.2 < M/{\rm M_\odot} <
5$.  With more sensitive spectropolarimetric studies,
\citet{auriere2010} showed that, at least for stars of spectral type~A
and late~B, there also appears to be a lower limit of $B_{\rm p}
\approx 300\,$G to the effective dipolar field strength, with no stars
observed with $30 < B_{\rm p}/{\rm G} < 300$.  Their lower limit is set
by sensitivity of their observations.  Thus, at least in this mass
range, there is evidence that main-sequence stars fall into two
groups, magnetic stars and non-magnetic stars.  We define the
poloidal magnetic flux $\Phi_{\rm p} = R^2 B_{\rm p}$, where $R$ is
the radius of the star.  Stars in this group are characterised by
\begin{equation}
10^{-8.5} < \frac{\Phi_{\rm p}/M}{\rm G\,cm^2\,g^{-1}} < 10^{-6.5},
\label{data}
\end{equation}
where $M$ is the total mass of the star.
The Magnetism in Massive Stars project \citep[MiMes,][]{wade2011} has
shown that a similar upper limit to the poloidal magnetic flux per
unit mass $\Phi_{\rm p}/M$ also applies to massive early-B and O~stars
\citep[see for example][]{wade2012}.  However it is not yet clear whether
the magnetic field dichotomy seen in the A~and late~B stars persists
at higher masses.

The origin of the large-scale magnetic fields observed in
main-sequence stars with radiative envelopes, namely the intermediate
mass main-sequence A, B~and O~stars, remains unresolved.  Whether the
fields are generated by contemporary dynamos in the stellar cores or
they are of fossil origin from pre-mainsequence phases of evolution
has been debated over the years.  A core dynamo is expected to
generate small-scale fields but the mechanism by which the field is
transported to the surface, where it is seen as a large scale ordered
field, has remained obscure.  Numerical magnetohydrodynamical
simulations \citep{braithwaite2009} of possible stable field
structures in stably stratified radiative stars appear to have swung
the balance in favour of the fossil field hypothesis.  These
calculations have confirmed that, while purely poloidal or purely
toroidal fields are subject to various instabilities that destroy the
fields on short time-scales, stability can be restored in combined
poloidal--toroidal field structures under certain conditions.  More
importantly, these calculations have shown that, regardless of the
nature, scale and complexity of the initial field, it relaxes to a
stable large-scale poloidal--toroidal structure on an Alfv\'en
crossing time-scale (about $10\,$yr for Ap~stars).  The subsequent
evolution of the field structure is expected to be slow enough for the
field to last for a time-scale similar to the main-sequence lifetime of
the star.  The presence of a long-lived, nearly dipolar field
structure in magnetic Ap~and Bp~stars is therefore seen to be
compatible with the fossil hypothesis.  What determines the maximum
flux per unit mass, however, remains unexplained but it must be
related to the origin of the magnetic flux acquired during the
pre-mainsequence history of the star.  The detection of magnetic
fields in some Herbig~Ae and Be~stars, with magnetic fluxes similar to
those seen in the strong field Ap~and Bp~stars \citep[for
  example][]{alecian2008,wade2007} suggests that the magnetic flux was
already present during the pre-mainsequence phase of evolution.

\citet{ferrario2009} suggested that magnetic main-sequence stars
acquire their magnetic fluxes through a dynamo process that occurs
during late merging of two protostars on their approach to the main
sequence.  Because only some stars have merged in such a way this
model can explain the magnetic dichotomy.  Here we investigate this
hypothesis further and suggest that any ordered primordial poloidal
field that is present in the merging stars can be amplified, by
differential rotation through a dynamo process, to a field
configuration which is mainly toroidal.  We present a simple model in
which magnetic fields decay whenever they are unstable to short
time-scale instabilities.  Otherwise toroidal field is generated at
the expense of differential rotation and poloidal field is generated
as a by product of the decay of the toroidal field.  Once differential
rotation has all been removed a long-lived magnetic configuration
remains.  We propose that such a dynamo, acting after stars merge,
provides an explanation for both the magnetic main-sequence stars and
the highest field magnetic white dwarfs.  The magnetic flux of
magnetars suggests that a similar dynamo may play a role in generation
of magnetic fields of at least some neutron stars.

\section {The Observed Maximum Poloidal Flux to Mass Ratio}

\begin{figure}
\begin{center}
\hspace{0.1in}
\epsfxsize=0.95\columnwidth
\epsfbox{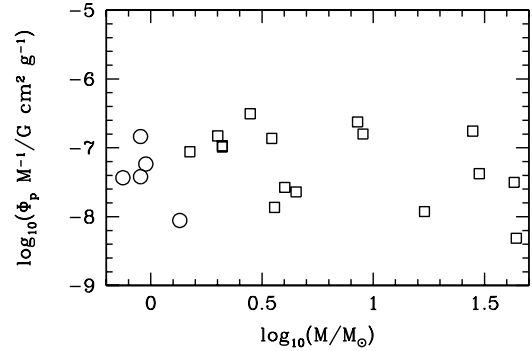}
\caption{The ratio of magnetic flux to mass $\Phi_{\rm p}/M$ for the
  most magnetic main-sequence stars (squares) and white dwarfs
  (circles).
\label{fluxes}}
\end{center}
\end{figure}

The incidence of magnetism among A~and B~stars peaks at about
$10\,$per cent at $3\,\rm M_\odot$ and decreases rapidly at lower
masses to effectively zero at $1.6\,\rm M_\odot$.  It appears possible
that the incidence increases further at higher masses
\citep{power2008} although this is yet to be established either way
through more complete surveys.  We note here that \citet{potter2012}
claim that there is an upper limit to the mass of a star that can
maintain magnetic field, for a significant fraction of its
main-sequence lifetime, in the presence of hydrodynamic instabilities.
The magnetic A, B~and O~stars have poloidal magnetic fluxes $\Phi_{\rm
  p} = B_{\rm p}R^2$ in a range that varies over many decades but they
all have a similar upper bound.  This is illustrated in
Fig.~\ref{fluxes} where we plot the poloidal magnetic flux per unit
mass, $\Phi_{\rm p}/M$, for the most magnetic main-sequence stars
\citep[][ and references therein]{wade2012}.

According to our hypothesis, the magnetic A, B~and O~stars all form
when proto-stellar objects merge late on their approach to the main
sequence.  Simulations of star formation have shown that the more
massive a star is, the more likely it formed by merging of lower mass
stars \citetext{Railton, Tout \& Aarseth, 2012, private
communication}.  So a consequence of our hypothesis is that the
incidence of magnetism, for as long as the field can be maintained,
increases with mass on the main sequence and there is some evidence
for this.  We have also noted \citep{ferrario2009} that such a merging
hypothesis provides a natural explanation for the apparent lack of Ap
stars in close binary systems with main-sequence companions.

Turning to white dwarfs we find that some $10\,$per cent have magnetic
field strengths in the range $10^6-10^9\,$G and form the group of
high-field magnetic white dwarfs \citep[HFMWDs,][]{wickramasinghe2005}.
We compare, in Fig.~\ref{fluxes}, the poloidal magnetic flux to mass
ratio of the highest field magnetic white dwarfs for which both field
and mass are known \citep{wickramasinghe2000} with those of the most
magnetic main-sequence stars.  The similarity in this ratio between
the two groups is striking and so suggests a common physical origin
for the magnetic fields.

Unlike for the main-sequence A stars, there is not such a clear
magnetic dichotomy in the white dwarfs.  Among the white dwarfs, there
is also a group of low-field magnetic white dwarfs (LFMWDs) with
fields ranging from a few~kG to $10^5\,$G or so.
\citet{landstreet2012} indicate that the discovery rate of white
dwarfs with fields in the range $10^3-10^5\,$G is significantly higher
than for the entire high field range of $10^6-10^9\,$G.  Combined with
the apparent dearth of fields in the range $10^5-10^6\,$G, this
suggests that the distribution may be bimodal.  Current surveys are
not geared to detect fields below 1\,kG so the possibility remains
that most white dwarfs are magnetic at some low level but this still
needs to be confirmed.  It seems likely that the magnetic fields in
the low-field group must have been generated through dynamo processes
in the course of normal single star evolution to the white dwarf
phase.

According to one school of thought a small proportion of single stars,
the magnetic main-sequence stars with strong fossil fields, may retain
their fossil flux through to the white dwarf phase to give rise to the
high HFMWDs \citep{wickramasinghe2005}.  This fossil from the main
sequence hypothesis, however, is not without difficulties.  The
progenitors of most white dwarfs have complex evolutionary histories
with different regions of the star becoming convective and radiative
as the star evolves through the red giant and asymptotic giant
branches.  The survival of fossil magnetic flux thus appears unlikely,
although this still needs to be established with detailed
calculations.  A more serious difficulty with this hypothesis is that
the birth rate from this route appears to be too small by a factor of
three or so to account for the entire class of HFMWDs \citep{Kawka2007}.
Alternative explanations for the fields seen in the HFMWDs have
therefore been sought.

From the absence of late-type star companions to HFMWDs, while such
stars are seen in abundance with other white dwarfs, \cite{tout2008}
established that high magnetic fields in white dwarfs are intimately
tied up with stellar duplicity.  They then proposed that magnetic
fields can be generated through differential rotation during the
common envelope evolution which leads to the formation of cataclysmic
variables.  Such fields could be trapped in the white dwarf and give
rise to the magnetic cataclysmic variables.  If the cores merge during
this phase, a highly magnetic isolated white dwarf can be formed.
Though the conditions are ripe for a strong dynamo in the common
envelope it is not easy to see how such a field could diffuse into the
white dwarf on the time for which the envelope remains in place
\citep{potter2010}.  We might envisage that a weak seed poloidal
field, generated when it was last convective and now anchored within
the white dwarf might be wound up by differential rotation in the
outer layers of the white dwarf during the common envelope evolution.
The ensuing dynamo action can enhance the poloidal field too as long
as both poloidal and toroidal fields are sufficient to stabilize one
another.  The final stable field structure can then freeze into the
outer layers of the white dwarf as it cools.

Some of the HFMWDs might have descended from double white dwarf
systems that merged, driven by gravitational radiation, after they
emerged from a common envelope in a close orbit.  Unlike those that
merged during the common envelope phase, which we expect to spin down
before ejection of their remaining envelope, these HFMWDs can still be
spinning more rapidly \citep{wickramasinghe2000}.  Even a small fraction of
circumstellar material accreted from a disc by the merged core can
spin its outer layers up to break up speeds \citep{guerrero2004} after
which a dynamo can generate the very strong magnetic fields.

\section{The Magnetic Flux to Mass Ratio and Stability of Field Structures}

For a star of mass $M$  and radius $R$, the thermal energy can be written as
\begin{equation}
U=\frac{\lambda GM^2}{R},
\end{equation}
where $G$ is the gravitational constant and $\lambda$ is a structural
constant of order unity. For a $2\,M_\odot$ A-star $\lambda\approx
0.5$.  The total magnetic energy can be written as
\begin{equation}
E = E_{\rm p} + E_\phi = \frac{4}{3}\pi R^3 (\frac{B_{\rm
p}^2+B_\phi^2}{2\mu_0}) = \frac{4}{3}\pi R^3\frac{B^2}{2\mu_0},
\end{equation}
where $B_{\rm p}$ is the poloidal, $B_\phi$ the toroidal and $B$ the
total magnetic field strength and
$\mu_0$ is the permeability of a vacuum.
Defining the ratios of magnetic to thermal energy by $\eta=E/U$,
$\eta_{\rm p}=E_{\rm p}/U$ and $\eta_\phi=E_\phi/U$,
then
\begin{equation}
\eta = \frac{B^2}{2\mu_0}\frac{4}{3}\pi R^3\frac{R}{\lambda GM^2} =
\frac{\Phi^2}{M^2}\frac{2\pi}{3\mu_0\lambda G}.
\end{equation}
Thence, scaling to the observed maximum poloidal flux, we find
\begin{equation}
\eta_{\rm p} = \frac{10^{-8}}{\lambda}\left(\frac{\Phi_{\rm p}/M}
{10^{-6.5}{\rm G\,cm^2\,g^{-1}}}\right)^2
\label{poloidalmax}
\end{equation}
and the observations demonstrate that the maximum $\eta_{\rm p}$ is
independent of the mass and type of star.

\citet{braithwaite2009} investigated the evolution of fossil fields
in non-rotating stars and found that for stable poloidal--toroidal
field structures, the energy in the toroidal and poloidal components
of the field must satisfy
\begin{equation}
a\eta^2 < \eta_{\rm p} < 0.8\eta,
\end{equation}
where $a$ is a buoyancy factor and $a\approx 10$ for main-sequence
stars.  The first inequality is due to the stabilizing effect of a
poloidal field on the Taylor instability of the $m=1$ mode in purely
toroidal fields.  A lower limit to the poloidal field is ultimately
set by the relative importance of magnetic to gravitational--thermal
energy through buoyancy effects.  The upper limit is due to the
stability of poloidal fields which requires they be not significantly
larger than the toroidal field.  These expressions were obtained
analytically from stability studies and were confirmed to be generally
consistent with results of numerical studies of non-rotating stars.
\cite{braithwaite2009} argued that the same inequalities are also
likely to hold for stable fields in rotating stars.

\section{The Dynamo Model}
\label{secdynamo}

In a similar spirit to \citet{tout1992} we set up a simple dynamo in
which toroidal field is generated from poloidal field by differential
rotation $\Delta\Omega$ within the star.  Unless the poloidal field is
strong enough to stabilize the toroidal the latter is lost by magnetic
instabilities which, in a non-rotating star, operate on an Alfv\'en crossing time-scale
\begin{equation}
\tau_{\rm A} = \frac{R\sqrt{\mu_0\rho}}{B},
\end{equation}
where $\rho = 3M/4\pi R^3$ is the star's mean density.
\citet{pitts1985} demonstrated that, in the presence of rotation, the
growth rate of magnetic instabilities is reduced by a factor
$\Omega/\Omega_B$, where $\Omega$ is the angular velocity and
$\Omega_B = 2\pi/\tau_{\rm A}$.
The observed spins of magnetic Ap stars are of the order $1$ to~$1{,}000\,$d
\citep{landstreet2000}, while their characteristic Alfv\'en
time-scale is of the order of~$2\,$d.  Similarly the observed spin rates of
magnetic white dwarfs are of the order of hours to years
\citep{wickramasinghe2000}, while their characteristic Alfv\'en
time-scale is about~$1.5\,$hr.  Therefore magnetic field decay is not
significantly limited by rotation once the magnetic fields have grown and
differential rotation removed.  However we begin our
calculations with small fields and differential rotation of the order
of the break up spin of the star.
The angular velocity at any point in the star lies between the
extremes of ineffectual, $\Omega < \Omega_{\rm B}$ and maximal,
$\Omega = \Omega_{\rm crit}$, where
\begin{equation}
\Omega_{\rm crit} = \sqrt{\frac{GM}{R^3}}
\end{equation}
is the surface break-up spin rate of the star.  While we envisage some
parts of the star to be rotating slowly others rotate faster by
$\Delta\Omega$.  Thus $0 < \Omega\approx\Delta\Omega < \Omega_{\rm
  crit}$ in regions where magnetic fields are decaying.
We therefore multiply the time-scale
for instability growth by $\Omega\tau_{\rm A}/2\pi$, where we first
set $\Omega = \Delta\Omega$ and later consider the two extreme cases
of $\Omega < \Omega_{\rm B}$ and $\Omega = \Omega_{\rm crit}$, when this is
longer than $\tau_{\rm A}$.  Our equation for the evolution of toroidal field is then
\begin{equation}
\frac{dB_\phi}{dt} = \Delta\Omega B_{\rm p} - \frac{B_\phi}{\tau_\phi},
\label{toroidal}
\end{equation}
where
\begin{equation}
\label{eqstabphi}
\tau_\phi =
\begin{cases}
\infty & \text{if $\eta_{\rm p} > a\eta^2$},\\
{\rm max}\left(1,\frac{\Omega\tau_{\rm A}}{2\pi}\right)\tau_{\rm
  A} & \text{otherwise}.
\end{cases}
\end{equation}
Unusually we regenerate poloidal field from the decaying toroidal
field with some efficiency $\alpha$.  Our reasoning is that decay of
toroidal field produces buoyant structures that generate a small scale
poloidal field.  Reconnection amongst these field elements leaves a
large scale poloidal field in a similar way to that described by
\citet{tout1996} for accretion discs.  In accord with
\citet{braithwaite2009} our poloidal field decays on the same modified
Alfv\'en time-scale unless there is sufficient total field strength to
stabilize it.  Thus
\begin{equation}
\frac{dB_{\rm p}}{dt} = \alpha\frac{B_\phi}{\tau_{\phi}} - \frac{B_{\rm
p}}{\tau_{\rm p}},
\label{poloidal}
\end{equation}
where
\begin{equation}
\label{eqstabp}
\tau_{\rm p} =
\begin{cases}
\infty & \text{if $0.8\eta > \eta_{\rm p}$},\\
{\rm max}\left(1,\frac{\Omega\tau_{\rm A}}{2\pi}\right)\tau_{\rm
  A} & \text{otherwise}.
\end{cases}
\end{equation}
Equations (\ref{toroidal}) and~(\ref{poloidal}) are non-linear because
of the dependence of $\tau_{\rm p}$ and~$\tau_\phi$ on $B$.
Differential rotation is gradually reduced by the magnetic torque so
that
\begin{equation}
I\frac{d\Delta\Omega}{dt} = -\frac{B_{\rm p}B_\phi}{\mu_0}4\pi R^2R.
\end{equation}
\par
We now write
\begin{equation}
\zeta = \sqrt{\eta} = B\sqrt{\frac{2\pi R^4}{3\mu_0\lambda G M^2}} =
B\sqrt{\frac{1}{2\mu_0\lambda}\frac{R}{GM\rho}},
\end{equation}
\begin{equation}
k = \frac{\Delta\Omega}{\Omega_{\rm crit}},
\end{equation}
and we define a dynamical time-scale
\begin{equation}
t_{\rm dyn} = \frac{1}{\Omega_{\rm crit}}.
\end{equation}
We introduce a dimensionless time
\begin{equation}
\tau = \frac{t}{t_{\rm dyn}}
\end{equation}
so that
\begin{equation}
\tau_{\rm A} = \frac{t_{\rm dyn}}{\sqrt{2\lambda}\zeta}
\end{equation}
and
\begin{equation}
\frac{\Delta\Omega\tau_{\rm A}}{2\pi} = \frac{k}{2\pi\sqrt{2\lambda}\zeta}.
\end{equation}
The moment of inertia of the star is
expressed as
\begin{equation}
I = \gamma MR^2,
\end{equation}
with $\gamma\approx 0.1$ for a $2\,M_\odot$ main-sequence star.
Our equations now take the dimensionless forms
\begin{equation}
\frac{d\zeta_\phi}{d\tau} = k\zeta_{\rm p} -
\sqrt{2\lambda}\sigma_\phi\zeta\zeta_{\phi},
\end{equation}
where
\begin{equation}
\sigma_\phi = \begin{cases}
0 & \text{if $\zeta_{\rm p}^2 > a\zeta^4$},\\
{\rm min}\left(1,\frac{2\pi\sqrt{2\lambda}\zeta}{k}\right)
& \text{otherwise},
\end{cases}
\end{equation}
\begin{equation}
\frac{d\zeta_{\rm p}}{d\tau} = \alpha \sqrt{2\lambda}\sigma_\phi\zeta\zeta_\phi -
\sqrt{2\lambda}\sigma_{\rm p}\zeta\zeta_{\rm p},
\end{equation}
where
\begin{equation}
\sigma_{\rm p} = \begin{cases}
0 & \text{if $\zeta_{\rm p}^2 < 0.8\zeta^2$},\\
{\rm min}\left(1,\frac{2\pi\sqrt{2\lambda}\zeta}{k}\right)
& \text{otherwise}
\end{cases}
\end{equation}
and
\begin{equation}
\frac{dk}{d\tau} = -\frac{6}{\gamma}\zeta_{\rm p}\zeta_\phi.
\end{equation}

\section{Results}

\begin{figure}
\begin{center}
\hspace{0.1in}
\epsfxsize=0.95\columnwidth
\epsfbox{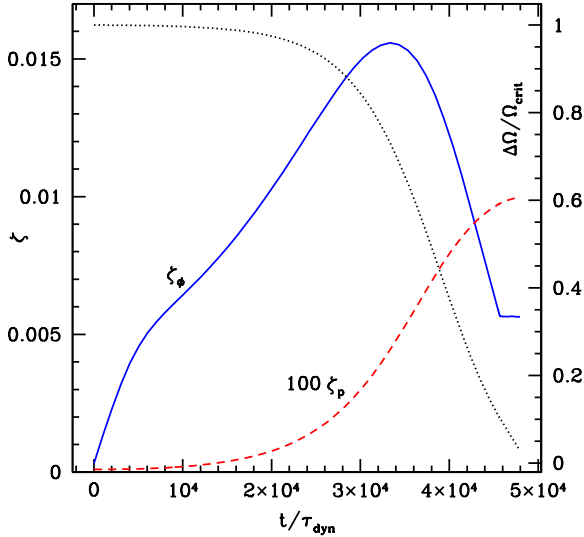}
\caption{The evolution of magnetic field components following a
collision that leaves the greatest differential rotation, equal to its
break up spin, in the merged object.  The decay of differential
rotation follows the right hand axis while the indicated fields
build up according to the left hand axis.  Toroidal field decays
unless $\eta_{\rm p} > a\eta^2$ and this determines the final ratio of
toroidal to poloidal field.  The poloidal field is somewhat weaker and
so is multiplied by $100$ on this figure.  Rotational stabilization of
magnetic field is included with $\Omega = \Delta\Omega$ in
equations (\ref{eqstabphi}) and~(\ref{eqstabp}).}
\label{figdynamo}
\end{center}
\end{figure}

We evolve the three coupled differential equations set up in
section~\ref{secdynamo} by a simple Euler method.  Experiments with
more precise and stabler algorithms revealed no perceptible
differences.  For a typical A-star of $2\,M_\odot$ the structural
constants $\lambda = 0.5$ and $\gamma =0.1$ are chosen according to
detailed models made with the Cambridge {\sc STARS} code
\citep{eggleton1971,pols1995} while $a = 10$ is chosen according to
\citet{braithwaite2009}.  This leaves a single free parameter
$\alpha$, which describes what fraction of the decaying toroidal field
is converted to poloidal, in our simple dynamo model.  For a given
initial amount of differential rotation we find that the final field
strengths increase monotonically with $\alpha$ because for higher
$\alpha$ less of the rotational energy is lost when the toroidal field
decays.  Thus we can calibrate $\alpha$ so that the final total
poloidal flux per unit mass is equal to that observed
(equation~\ref{poloidalmax}).  Fig.~\ref{figdynamo} shows the
evolution of magnetic fields as the differential rotation decays away
from its maximum, $\Delta\Omega = \Omega_{\rm crit}$.  We have set
$\alpha = 1.52\times 10^{-4}$ to reproduce the observed maximum in
$\eta_{\rm p}\approx 10^{-8}$, so $\zeta_{\rm p} \approx 10^{-4}$
again with $\lambda = 0.5$.  For small enough seed fields toroidal
field is initially generated by winding poloidal field owing to the
differential rotation.  As soon as the toroidal field is large enough
that stability criterion~(\ref{eqstabphi}) is violated the poloidal
field begins to grow.  While differential rotation is still sufficient
the toroidal field grows but it eventually reaches a maximum and
decays until it reaches equilibrium with the poloidal field so that
$a\eta^2 = \eta_{\rm p}$.  Throughout this evolution $\Delta\Omega$ is
destroyed by the magnetic torque so that the final object is rotating
as a solid body.  At this point $\tau = 4.88\times 10^4$.  The
dynamical time-scale for a $2\,M_\odot$ star is about 40\,min so this
evolution of the magnetic fields is over in about 3.7\,yr which is much
less than the corresponding Kelvin-Helmholtz time-scale of $2.3\times
10^6\,$yr for a $2\,M_\odot$ star.  The field configuration is stable
according to \citet{braithwaite2009}'s criteria and hence can survive
until thermal or nuclear time-scale evolution destroys it.

\begin{table*}
\caption{Dynamo models varying initial conditions and the parameter
$\alpha$.  Columns are the fraction $\alpha$ of decaying toroidal
field that becomes poloidal, the initial ratio of differential
rotation to break up spin, the initial and final ($\Delta\Omega =
0$) toroidal and poloidal fields and the number of dynamical times to
reach to reduce $\Delta\Omega$ to zero.  The first row shows the model
which best fits the observed maximum total flux per unit mass and
which is illustrated in Fig.~\ref{figdynamo}.
\label{tabledynamo}}
\begin{tabular}{@{}rcccccc}
\hline
$\alpha$ & $k_{\rm i}$ & $\zeta_{\phi,{\rm i}}$ & $\zeta_{\rm p,i}$ & $\zeta_{\phi,{\rm f}}$ & $\zeta_{\rm p,f}$ & $\tau_{\rm f}$ \\
\hline
$1.52\times 10^{-4}$ & $1  $ & $10^{-3.5}$ & $10^{-6}$ & $5.36\times 10^{-3}$ & $           10^{-4}$ & $4.88\times 10^4$ \\ 
$1.52\times 10^{-4}$ & $1  $ & $10^{-4.5}$ & $10^{-6}$ & $5.35\times 10^{-3}$ & $           10^{-4}$ & $4.91\times 10^4$ \\ 
$1.52\times 10^{-4}$ & $1  $ & $10^{-2.5}$ & $10^{-6}$ & $5.35\times 10^{-3}$ & $           10^{-4}$ & $4.60\times 10^4$ \\ 
$1.52\times 10^{-4}$ & $1  $ & $10^{-3.5}$ & $10^{-7}$ & $5.31\times 10^{-3}$ & $9.99\times 10^{-5}$ & $8.41\times 10^4$ \\ 
$1.52\times 10^{-4}$ & $1  $ & $10^{-3.5}$ & $10^{-5}$ & $5.45\times 10^{-3}$ & $1.03\times 10^{-4}$ & $2.74\times 10^4$ \\ 
$1.5 \times 10^{-3}$ & $1  $ & $10^{-3.5}$ & $10^{-6}$ & $1.24\times 10^{-2}$ & $5.55\times 10^{-4}$ & $1.08\times 10^4$ \\ 
$1.5 \times 10^{-5}$ & $1  $ & $10^{-3.5}$ & $10^{-6}$ & $2.31\times 10^{-3}$ & $1.79\times 10^{-5}$ & $3.22\times 10^5$ \\ 
$1.15\times 10^{-4}$ & $0.5$ & $10^{-3.5}$ & $10^{-6}$ & $3.94\times 10^{-3}$ & $4.94\times 10^{-5}$ & $8.08\times 10^4$ \\ 
$1.15\times 10^{-4}$ & $0.1$ & $10^{-3.5}$ & $10^{-6}$ & $1.67\times 10^{-3}$ & $1.67\times 10^{-6}$ & $2.29\times 10^5$ \\ 
\hline
\end{tabular}
\end{table*}

Table~\ref{tabledynamo} lists the final state for various initial
conditions and variation of the regeneration parameter $\alpha$.  The
first row is the model of Fig.~\ref{figdynamo} already discussed.  The
next two rows illustrate how changing the seed toroidal field has
almost no effect.  Its growth or decay rapidly wipe out any initial
state.  Rows four and five show that varying the initial seed poloidal
field also has little effect on the final magnetic fields.  However
reducing the seed poloidal field does mean that the differential
rotation lasts longer.  This is simply because the initial generation
of toroidal field depends on the presence of poloidal field.  Once the
poloidal field has built up the evolution speeds up and proceeds in a
similar way to much the same end point.  Rows six and seven
demonstrate how the parameter $\alpha$ influences the final field
strengths.  For large $\alpha$ more of the energy extracted from
differential rotation goes into magnetic fields.  If $\alpha$ were
unity no energy would be lost until the stability
criterion~(\ref{eqstabp}) were violated.  We do not expect the
generation of poloidal field by decaying toroidal field to ever be
sufficient to violate this criterion once the growth of the fields
from their seeds is established.

The last two rows of Table~\ref{tabledynamo} show how the final
magnetic flux depends on the initial differential rotation.  The final
poloidal flux is almost directly proportional to the initial $k_{\rm i}$.
Thus we hypothesize that, in collisions, the differential rotation
left within the merged object ranges up to the maximum which cannot be
higher than the break up spin.  The observed lower limit to the total
flux per unit mass in main-sequence stars would then result from a
lower limit, of about 1\,per cent of the break up spin, to the
differential rotation within merged main-sequence stars.  It is
reasonable to imagine that stars are unlikely to merge without some
residual angular momentum because head on collisions are extremely
rare.  In the case of white dwarfs that merge during common envelope
evolution we might then expect a similar minimum magnetic field of
some $10^7\,$G compared with the maximum of $10^9\,$G observed for
single white dwarfs.  Here the observations suggest a somewhat lower
value of about $10^6\,$G as a lower field limit for the HFMWDs
\citep{Kulebi2009}.

\begin{figure}
\begin{center}
\hspace{0.1in}
\epsfxsize=0.95\columnwidth
\epsfbox{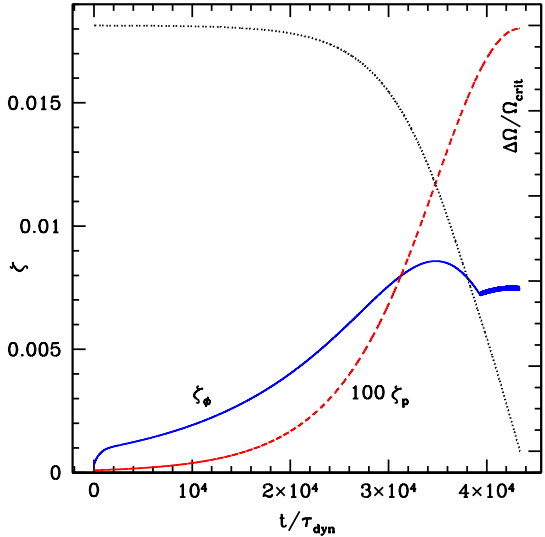}
\caption{The dynamo when 
rotational stabilization of field decay is not included, $\Omega <
\Omega_{\rm B}$ throughout.  Initial conditions and $\alpha$ are the
same as for the model of Fig.~\ref{figdynamo}.  In this case toroidal
field decay and consequent poloidal field growth begin much earlier so
that the final poloidal field is larger by a factor of 1.8.}
\label{figdynamoomegazero}
\end{center}
\end{figure}

\begin{figure}
\begin{center}
\hspace{0.1in}
\epsfxsize=0.95\columnwidth
\epsfbox{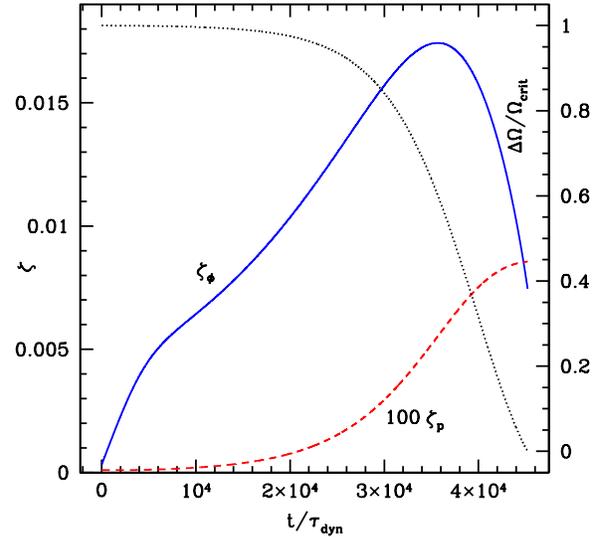}
\caption{The dynamo when field decay is maximally rotationally
stabilized, $\Omega = \Omega_{\rm crit}$ throughout.  Initial conditions and $\alpha$ are the
same as for the model of Fig.~\ref{figdynamo}.  Early on
$\Delta\Omega\approx\Omega_{\rm crit}$ so the evolution is very
similar.  At late stages toroidal field decay is more stabilized so
that the final toroidal field is slightly larger and the final
poloidal field is smaller by a factor of 0.85.}
\label{figdynamoomegacrit}
\end{center}
\end{figure}

To investigate the importance of rotational stabilization of magnetic
field decay \citep{pitts1985} we consider the two extreme cases of
minimal and maximal rotational stabilization.  All observed high field
stars now have $\Omega < \Omega_{\rm B}$ at their surfaces.
Fig.~\ref{figdynamoomegazero} shows the evolution from the same
initial conditions and with the same field generation parameter
$\alpha$ as the model of Fig.~\ref{figdynamo} but with $\Omega <
\Omega_{\rm B}$ throughout the evolution.  As in that case the
toroidal field initially builds up at the expense of differential
rotation but decay sets in earlier leading the generation of more
poloidal field throughout the evolution.  At the other extreme we set
$\Omega = \Omega_{\rm crit}$ throughout the evolution.  In this case,
shown in Fig.~\ref{figdynamoomegacrit} for the same initial conditions
and $\alpha$, the early evolution is not much altered because
$\Delta\Omega\approx\Omega_{\rm crit}$.  Once $\Delta\Omega$ falls
toroidal field builds up more at the expense of generating poloidal so
that the final poloidal field is smaller.  However we note that the
differences are not large and so our model is quite robust to the
choice of $\Omega$.  The final poloidal field is in the range given by
$8.5\times 10^{-5} < \zeta_{\rm p} < 1.8\times 10^{-4}$ from the
extreme of maximal rotational stabilization to none at all.

Single HFMWDs can form either when two cores merge during common
envelope evolution or when two white dwarfs of total mass below the
Chandrasekhar limit emerge from a common envelope with a short enough
period and mass ratio close enough to unity that they subsequently
merge owing to angular momentum loss by gravitational radiation.
\citet{zhang2009} plot the distribution of magnetic fields of single
HFMWDs and magnetic cataclysmic variables in their fig.~5.  The single
star fields follow much the same distribution as the combined polar
and intermediate polar distribution up to $3\times 10^8\,$G or so.
Higher fields are found only in the single white dwarfs.  In a common
envelope, a low-mass main-sequence star merges when it has spiralled
in enough to be tidally disrupted by the degenerate core and it
dissolves into the envelope.  \citet{tout2008} hypothesized that the
magnetic cataclysmic variables must have emerged from their common
envelopes only just detached in order to have acquired their strong
fields.  
For these we envisage that
the hot outer layers of the degenerate core are weakly spun up by the
differential rotation between the orbit of the cores and the envelope.
A dynamo of the sort discussed here could then build up strong fields
from a weak seed field in much less time than would be required for
magnetic fields to diffuse into the white dwarf \citep{potter2010}.
This is much more consistent with the expected time-scale for common
envelope evolution.
Those that merge account for the bulk of the single HFMWDs that have a
similar distribution of fields to the magnetic cataclysmic variables.
White dwarfs in the high-field tail have then come from merging
degenerate objects.  We expect those that merge during common envelope
evolution to spin down considerably before the remainder of their
envelope is ejected while those that merge subsequently may continue
to spin more rapidly as solid bodies.  While most magnetic white
dwarfs have spin periods far in excess of $1\,$d a few spin more
rapidly.  For example RE\,J0317--853 has a spin period of only
$725\,$s \citep{barstow1995} and a very high magnetic field of $450\,$MG
\citep{ferrario1997}.  We would expect that the highest fields are, on
average, associated with higher mass white dwarfs.  Future population
synthesis studies coupled with more extensive observed data will test
this hypothesis further.

Lastly we comment on the possible relevance of our model to neutron
star magnetic fields.  Among the neutron stars are the magnetars
characterised by their extremely strong magnetic fields of
$10^{14}-10^{15}\,$G \citep{rea2011}.  In a sense these are the
counterparts of the HFMWDs.  Merging of two white dwarfs with a total
mass that exceeds the Chandrasekhar mass could lead to a collapse to a
neutron star under certain circumstances.  The white dwarf type fields
generated during such a merge could be amplified to magnetar-strength
fields by flux conservation during the subsequent collapse
\citep{King2001}.  However, given the large fraction of magnetars among
neutron stars, the majority of them are likely to have formed as end
products of single star evolution.  Prior to collapse, the cores may
be expected to have a range of spin periods depending on the initial
angular momenta of stars on the main sequence and angular momentum
transport during stellar evolution \citep{Heger2005}.  Models predict
that, if angular momentum is conserved during the subsequent core
collapse, the resulting neutron stars have spin periods that are
generally in accord with the birth spins observed for pulsars.  If
strong differential rotation is acquired during core collapse,
magnetic fields may be generated by a dynamo of the type that we have
proposed.  We estimate that poloidal fields of $2\times 10^{14}-1.2
\times 10^{15}\,$G, similar to magnetar fields, could be generated in
a neutron star that, at birth, differentially rotates at $5-30\,$per
cent of its break up speed and if the dynamo efficiency factor
$\alpha$ is similar to that which we deduced for main-sequence stars.
Such magnetars would also harbour even stronger toroidal fields and we
note, significantly, that most models of magnetars require such fields
to explain their emission properties \citep{rea2011}.

\section {Conclusions}

Observations of magnetic main-sequence stars and white dwarfs
show that there appears to be a universal upper limit to the poloidal
magnetic flux per unit mass.  We have explored recent proposals that
the strongly magnetic stars in both these groups arise when stars
merge.  A simple magnetic dynamo that extracts energy from
differential rotation explains the limit if the maximum fields are
generated in merged objects that are differentially rotating at
breakup.  Weaker fields are generated in merged objects with less
differential rotation.

In the case of white dwarfs we envisage that common envelope evolution
generates differential rotation in the hot outer parts of the
degenerate core.  For cataclysmic variables the amount of differential
rotation that can be imparted depends on how close the orbiting cores
get before the envelope is lost.  A dynamo such as we have described
can operate on a rapid time-scale to create stable fields that become
frozen in to the white dwarf as it cools.  When cores merge in a
common envelope we do not expect a collision.  Rather a main-sequence
star would dissolve on reaching a depth in the envelope at which it is
tidally disrupted by the degenerate core.  The fields generated ought
to be similar to those in the magnetic cataclysmic variables, which we
envisage to have ejected their common envelopes just before they would
otherwise have merged.  Similarly a lower mass degenerate core would
be tidally shredded and then accrete on to the more massive degenerate
core either within the common envelope or subsequently if a double
degenerate system merges by gravitational-wave angular momentum loss.
In both cases we expect the differential rotation that can be built up
in the outer layers of the the hot or heated core to be very large
because accretion of less than a tenth of a solar mass of material
from the inner edge of an accretion disc can spin a substantial
fraction of the white dwarf to break up.  So the maximum magnetic
field strengths should be found only in the single white dwarfs formed
from merged degenerate cores or stars.  Those that merge within the
common envelope we expect to rotate very slowly, spun down as angular
momentum is lost with the remaining envelope.  Those that merge
subsequently to the envelope ejection can be expected to have a high
fields too but also to be more rapidly rotating.

Merging of two white dwarfs with a total mass that exceeds the
Chandrasekhar limit may under certain circumstances lead to an
accretion induced collapse to a neutron star with magnetar type
fields.  Though most magnetars are likely to have evolved from single
stars, a similar dynamo may still play a role to generate
fields in these stars if they form highly differentially rotating
after core collapse.

\section*{Acknowledgments}

We thank the referee for drawing out attention to the work by
\citet{pitts1985}.  DTW and LF thank the Institute of Astronomy for
hospitality while this work was continued.  CAT thanks the University
of Cambridge, Monash University and the Australian National University
for supporting a visit to Australia during which this work was begun,
Churchill College for his fellowship and the Bombay Cambridge Society
and the Inter University Centre for Astronomy and Astrophysics in Pune
for providing an atmosphere in which the writing could be completed.

\label{lastpage}

\end{document}